\documentclass[fleqn,usenatbib]{mnras}

\usepackage{newtxtext,newtxmath}

\usepackage[T1]{fontenc}
\usepackage{ae,aecompl}

\usepackage{graphicx}	
\usepackage{amsmath}	
\usepackage{amssymb}	

\usepackage{csquotes}	
\usepackage{array}      
\usepackage{makecell}   

\title[NGC 5273 X-ray variability ]{X-ray reverberation lags from the 1.5 Seyfert galaxy \\ NGC 5273 }

\author[Vincentelli et al.]{
	F. M. Vincentelli$^{1}$ \thanks{E-mail: F.M.Vincentelli@soton.ac.uk } ,G. Mastroserio $^2$, I. Mc Hardy$^{1}$, A. Ingram $^3$, M. Pahari $^1$\\
	$^{1}$Department of Physics \& Astronomy, University of Southampton, Highfield, Southampton SO17 1BJ, UK\\
	$^{2}$Astronomical Institute, Anton Pannekoek, Univerity of Amsterdam, Science Park 904, NL-1098 XH Amsterdam, Netherlands\\
	$^3$Department of Physics, Astrophysics, University of Oxford, Denys Wilkinson Building, Keble Road, Oxford, OX1 3RH, UK
}
\date{Accepted XXX. Received YYY; in original form ZZZ}

\pubyear{2015}

\begin{document}
	\label{firstpage}
	\pagerange{\pageref{firstpage}--\pageref{lastpage}}
	\maketitle
	
	\begin{abstract}
		
		We present the results of X-ray spectral-timing analysis of a 90ks XMM-Newton observation of the nearby, broad line, early-type AGN NGC5273.  The X-ray spectrum revealed the clear presence of a reflection component at high energies, with a clear signature of a narrow iron line at 6.4 keV, consistent with distant reflection. Applying a relativistic reflection model, we found only marginal evidence for a broader relativistic line component.  However, cross-spectral analysis revealed that, between 4 and 6 $\times 10^{-4}$ Hz, the 5-8 keV band lagged the 2-3 keV band, implying reflection of the iron line from material close to the black hole. From the analysis of the lag-energy spectrum, we found a broad, but skewed line with a peak of $\approx$ 1000s at 7.5 keV relative to the continuum, which we interpret as the iron line in the reverberation spectrum from an illuminated accretion disc.  From the asymmetry in the shape of lag-energy spectrum, we also found that the source is consistent with having an inclination $\geq 45^\circ$.\end{abstract}
	
	
	

	Active Galactic Nuclei (AGN) are believed to originate from accretion of matter onto super-massive black holes (10$^6$-10$^9$ M$_\odot$) at the centers of galaxies. They produce emission from radio up to gamma-rays, with a complex spectrum deriving from a number of different physical components \citep{fabian2012,padovani2017}. Strong outflows, and the copious amounts of radiation produced by these objects are observed to  affect the properties of their host galaxies \citep{Schawinski2007,cicone2014,king2015}. 
	Most of our understanding, however, is still limited by large uncertainties concerning the precise geometry of these systems. Constraining physical parameters such as black-hole mass and disc inclination is therefore crucial to understand the underlying physical processes and their consequences on larger scales \citep[see e.g.][]{bentz2014,graham2011}.
	
	Historically, the study of the X-ray emission and its variability have proven to be among the best approaches for shedding light on the physical processes taking place in the innermost regions, close to central compact object. The primary X-ray radiation is thought to be produced in a hot electron corona by Compton up-scattering lower energy disc photons \citep{haardt_maraschi1991}. This radiation illuminates the surrounding  relatively cold accretion disc \citep{guilbert_rees1988,fabian1989}, giving rise to a characteristic reflection spectrum \citep{ross-fabian1993,garcia2013}. One of the most prominent features of this reflection spectrum is the presence of the iron K$\alpha$ line at $6.4$ keV.  In type-1 Seyfert galaxies this feature was observed to be broadened and skewed, which was successfully explained in terms of general relativistic effects in the strong
	gravitational field near to the central black hole \citep{tanaka1995}.

	X-ray variability is known to be a fundamental property of AGNs which has
	enabled the extraction of key information on the physical properties of these systems \citep{uttley2001,uttley2005,vaughan2003}. The timescales on which the high energy emission of these sources can vary span from  years down to $\approx$ 100 s \citep{lawerence1987,McHardy1987}. The Fourier power spectral density (PSD) measured from the emission of these sources is characterized by a decreasing broken power-law trend (e.g. the slope increases from $\approx -1$ at lower frequencies to $\approx -2$ at higher frequencies) with a break frequency which scales with the mass \citep{uttley2002,mchardy2004,mchardy2006}.

	Recent long and continuous exposures performed by XMM-Newton have permitted detailed study of the X-ray cross-spectral properties of AGN, down to $\approx$ 0.5 keV, on very short timescales. The main result from this approach has been  the discovery of a lag between the X-ray continuum and the radiation emitted both in the iron emission lines and in the low energy (<2 keV) reflection continuum.  In particular \citet{fabian2009} produced a firm detection in 1H 0707-495 that the variability in the Iron L line (0.3-1 keV) lagged the rest of the continuum. Marginal evidence for a soft lag in the same energy range had already been found by \citealt{markowitz2007,mchardy2007} for Mkn 766 and Ark 564 respectively.  Further observations revealed the presence of such a lag in other sources both in the low energy continuum and also in the K$\alpha$ line \citep{Emmanoulopoulos2011,zoghbi2011,zoghbi2012,zogbhi2013,kara2013a,kara2013b}.  
	A correlation between the mass and the amplitude of the lag among several Seyfert type-1 galaxies \citep{demarco2013,kara2016} was discovered, suggesting a common geometry among these sources.
	These results motivated more quantitative modeling of the reverberation phenomenon. Initially, the work focused on reproducing the Fourier-resolved lags between two X-ray energy bands using analytic response functions for different geometries \citep{Emmanoulopoulos2011, alston2014}. 
	More detailed work \citep[e.g.][]{Emmanoulopoulos2014, cackett2014} included response functions based on numerical General Relativistic ray tracing to determine black hole mass, spin, inclination and X-ray source height assuming the lampost model.
	More recently, it has become possible to also include the energy dependence of the lags in the models.  This has allowed
	the properties of the accretion process to be investigated more comprehensively, including additional parameters such as the shape of the X-ray source (lamp-post or extended corona)  \citealt{cackett2014,wilkins2016,chainakun2017,caballerogarcia2018,taylor2018,chainakun2019,ingram2019}.

	NGC 5273 is a nearby  (z=0.00362) \citep{bentz2014} low luminosity AGN (LLAGN) hosted by a lenticular galaxy (S0 morphology) at a distance of 16. $\pm$ 1.6 Mpc \citep{tonry2001}. First known as a 1.9  Seyfert Galaxy \citep{osterbock}, it has recently been re-classifed as 1.5 Seyfert galaxy after
	a re-analysis of the contribution of the host galaxy to its spectrum revealed the presence of broad components to  $H_\alpha$ and $H_\beta$ \citep{trippe2010}.  
	Through optical reverberation mapping measurements, \citet{bentz2014} measured a black hole mass of  $4.7 \pm 1.1 \times 10^6 M_{\odot}$.  
	
	X-ray observations  with Suzaku showed that NGC~5273 was significantly variable in X-rays \citep{kawamuro2016}. Recent analysis of a {Swift+NuSTAR} observation by \citet{pahari2017} found  clear evidence for non-relativistic reflection in the X-ray spectrum, with also a marginal detection of a high energy cut-off at $140$ keV.  In this work we analyse a 90 ks XMM-Newton X-ray observation, confirming rapid short timescale variability (Section 2). In Section 3 we analyse the time averaged X-ray spectrum under various model assumptions, looking also at the X-ray inter-band lags. We then  discuss the implications of these results for the source geometry in Section 4.  
	\section{Data Reduction}

\begin{figure}
		\centering
		\includegraphics[width=\columnwidth]{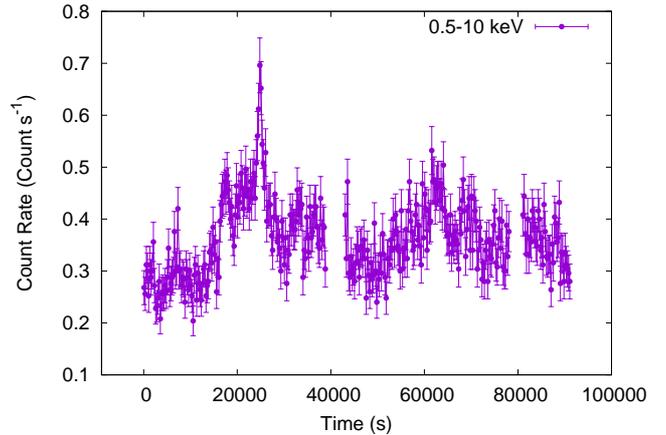}
		\caption{X-ray (0.5-10 keV) light curve extracted using the  EPIC-pn data. Time resolution is of 100 s.  }
		\label{fig:lcurve}
	\end{figure}

	Data analysed in this work were taken in {\it Full Frame} mode with the EPIC-pn camera on board of the {\it XMM-Newton} satellite during a 90 ks pointing performed on 02/06/2017 (OBSID 0805080401). 
	Events were extracted within a circle of $45$ arcseconds radius around the source and filtering events with PATTERN $<=4$ and FLAG==0 . Events were then binned in a lightcurve with a time resolution of $100$ s (see Fig.~\ref{fig:lcurve}). The source was found to vary significantly within the observation, with a mean count rate in the energy range  $0.5-10$ keV of $0.361 \pm 0.003$ counts s$^{-1}$ and a fractional rms of $0.16$. 
	In order to quantify the background contribution, events were extracted from a clear circular region in the field with the same aperture size as the source region. Two small segments (each of duration $3$ ks) in which the background count rate exceeded $0.04$ counts s$^{-1}$  were excluded from the analysis. 
	The same two source and background regions were used to extract the time-averaged energy spectrum. The energy spectrum was then grouped in order to have a minimum of 20 counts per energy bin.
	On top of the statistical errors we added $1\%$ systematic errors when  fitting the time-averaged energy spectrum.

	\section{Data Analysis}

	\subsection{Spectral Analysis}
\begin{figure}
		\includegraphics[width=\columnwidth]{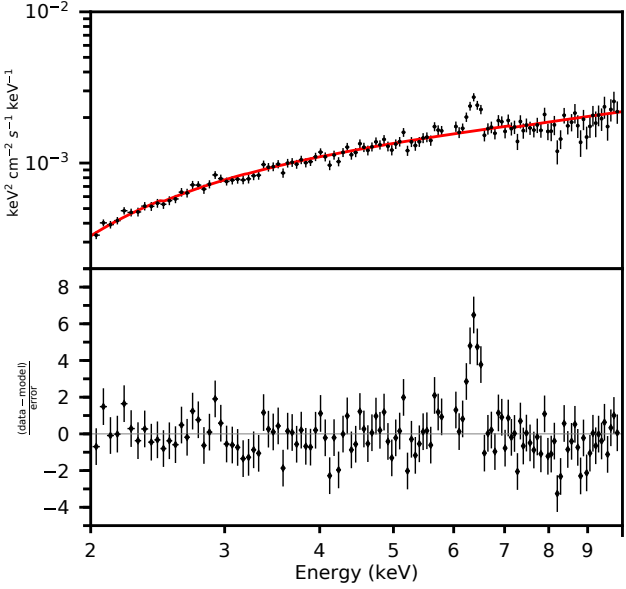}
		\caption{Unfolded spectrum (upper panel) and the residuals (bottom panel) of the best fit using only an absorbed power-law in the energy range $2-10$ keV. An excess at 6.5 keV is evident.}
		\label{fig:ratio}
	\end{figure}

\begin{figure*}
		\includegraphics[width=\textwidth]{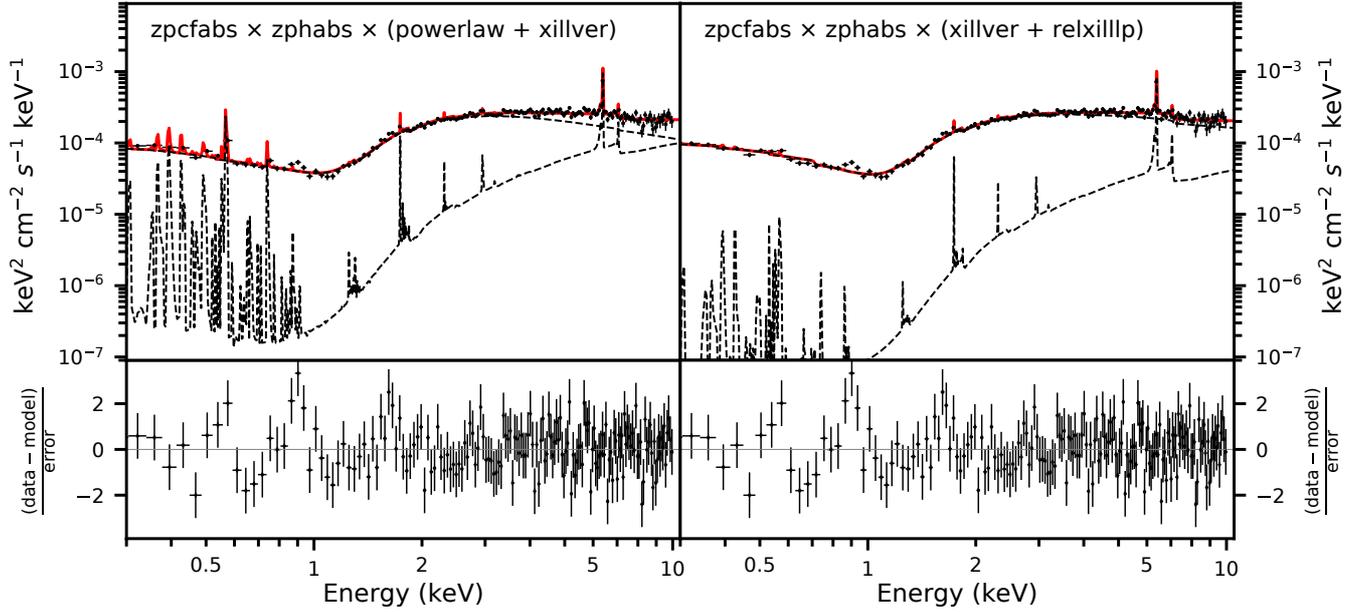}
		\caption{Unfolded spectra with the best fitting models (red line) and its residuals using an absorbed power-law and \textsc{xillver} (left panels) and using \textsc{relxilllp} + \textsc{xillver} (right panels) in the energy range $0.3-10$ keV. The components of each models are listed in the plots and represented with dashed lines.}
		\label{fig:final_spec}
	\end{figure*}

	The time-averaged energy spectrum is strongly absorbed at energies below $2$ keV. We therefore first considered the time-averaged spectrum between $2$ and $10$ keV. Fig~\ref{fig:ratio} shows the residuals of fitting an absorbed power-law (\textsc{tbabs + powerlaw}), with the Galactic hydrogen column density fixed to $n_{\rm H} = 9.2 \times 10^{19}$ cm$^{-2}$ (\citealt{pahari2017}). All the spectral models reported throughout this paper use this same fixed hydrogen column density to  account for  absorption in our galaxy. The residuals show a significant \textit{narrow} excess between $6$ and $7$ keV (Fig. \ref{fig:ratio}), suggesting the presence of a non-relativistic, distant reflection component (see also \citealt{pahari2017}).  We modeled this feature using the \textsc{xillver} model (\citealt{garcia2013}) and using the power-law model for the inverse-Compton emission. We then also included the softer $0.3 - 2$ keV energy range.  
	The soft part of the time-averaged spectrum shows strong absorption due to gas of the host galaxy  in our the line of sight. 
	We modeled this extra absorption component with the partial covering model \textsc{zpcabs}
	which accounts for the cosmological redshift of the galaxy ($z = 0.00362$).  
	Following the procedure described in \citet{pahari2017}, we fixed the inclination (i=40$^\circ$) 
	because the spectral fit is not very sensitive to this parameter.  
	The illuminating power-law index of \textsc{xillver} is fixed to the power-law index representing the direct emission, and the same procedure is applied to the high energy cut-off, which is fixed to $140$ keV (\citealt{pahari2017}) because the energy range of XMM-Newton is too low to put a reliable constraint on this parameter.  Top panel in Fig~\ref{fig:final_spec} shows the unfolded spectrum with the best fitting model (red line), with the different components (dotted lines).
	The bottom panels show the residuals of the fit. 
	Addition of the reflection component   improves the residuals structure in the iron line energy range and leads to an acceptable reduced $\chi^2$ of $180.5/163$. 
	
	We measured a total flux between 0.5 and 10 keV of 3.3$^{+0.01}_{-0.1} \times 10^{-12}$ erg cm$^{-2}$ s$^{-1}$;  in the 0.5--2 and 2--10 keV  bands we measured respectively 1.9 $^{+0.02}_{-0.06} \times 10^{-13}$ erg cm$^{-2}$ s$^{-1}$ and  3.1$^{+0.03}_{-0.1} \times 10^{-12}$ erg cm$^{-2}$ s$^{-1}$.
	The second column of Table~\ref{tab:parameters} reports the parameter values of the model. 
	In order to model the strong soft absorption, the partial covering model (\textsc{zpcf}) has a very high covering factor ($f_c= 0.958^{+0.003}_{-0.006} $) with 
	a hydrogen column density of $N_H=2.51^{+0.05}_{-0.05}\times 10^{22}$ cm$^{-2}$.
	We note that the residuals present some excesses for energies lower than $2$ keV, in particular apparent absorption features seem to be present at $\approx 0.7$  and $\approx 1.3$  keV. 
	Similar features were already seen in the highly absorbed type 2 Seyfert Galaxy IRAS 18325-5926 \citep{lobban2014}, {but may also be due to variable abundances in addition to iron.  A more detailed modelling, however, is beyond the scope of this paper.}
	The index of the illuminating power-law is $\Gamma=1.86^{+0.01}_{-0.01} $ which is very similar to the value \citet{pahari2017} found with Swift/XRT and NuSTAR ($\Gamma=1.81^{+0.02}_{-0.03} $). 
	The iron abundance of the reflection model is pegged to the lowest allowed value of \textsc{xillver} ($0.5$), whereas it is close to unity in the model of \citet{pahari2017} ($ 1.2 \pm 0.3 $). 
	The distant reflecting plasma seems to have a very low ionization ($\log\xi = 0.33^{+0.04}_{-0.03}$), 
	as expected far from the black hole.  
	{As a further test we also re-performed the fit leaving free the high-energy cut-off. However, no significant variation was found.}
	
	Many AGNs show an additional {\it broad} relativistic reflection component  around the iron $K\alpha$ energy in their time-averaged  spectra.  
	We therefore also used the \textsc{relxillp} model (\citealt{Dauser2014,Garcia2014})  which accounts for both the direct and the  reflected emission. 
	The former is described with a cut-off power-law and the latter is the radiation emitted from a lamp-post source above the black hole and reflected by a geometrically thin but optically thick accretion disc. We consider a constant radial ionization profile in the accretion disc. 
	The parameters shared by the distant reflection and the relativistic reflection emission such as the power-law index of the illuminating radiation, its high energy cut-off, the iron abundance in the disc and the inclination of the system are tied together. Except for the high energy cut-off and the black hole spin ($a$), these parameters are free in the fit, as are the height of the source ($h$), the inner radius of the disc and the reflection fraction (refl\_frac). {Even though black hole spin is known to affect the reflection spectrum \citep[see e.g.][]{martocchia2000}, the model is more sensitive to the inner disc radius than the actual spin value \citep{dauser2013}. Here we are more interested in the extension of the accretion disc, therefore we consider a maximally spinning black hole with a free inner radius parameter. Thus we allow for the possibility that the accretion disc does not reach in as far as the ISCO.}
	The distant reflector is considered neutral so its ionisation parameter is set to $0$.
	Table~\ref{tab:parameters} shows the values of the parameters for the best fitting model.
	We note that using two refection components (non-relativistic and relativistic reflection) improves the reduced $\chi^2$.
	However, both models are statistically acceptable and the F-test shows that the relativistic component 
	is significant with $2\sigma$ confidence.

	\begin{table}
		\caption{Best fit parameter values for time-averaged energy spectrum. The models considered in the fit are specified in the table. The hydrogen column density for the galactic absorption, the high energy cut-off and the spin of the black hole are fixed to $0.0092 \times 10^{22}$ cm$^{-2}$, $140$ keV and $0.998$ respectively. The errors are reported at $90\%$ confidence level.}
		\renewcommand{\arraystretch}{1.5}
		\begin{tabular}{ p{2.8cm} | p{2.8cm}cc p{2.8cm}cc }
			\hline
			
		\end{tabular}
		\begin{list}{}{}
			\item[$^a$] In the \textsc{xillver}+\textsc{relxill} model \textsc{xillver}'s ionisation is $0.0$  
			\item[$^b$] The lower limit of the iron abundance is $0.5$
			\item[$^c$] The lower limit is at ISCO. 
		\end{list}
		\label{tab:parameters}
	\end{table}
	
	\subsection{Long term spectral evolution}
	
\begin{figure} 
		\centering
		\includegraphics[width=0.5\textwidth]{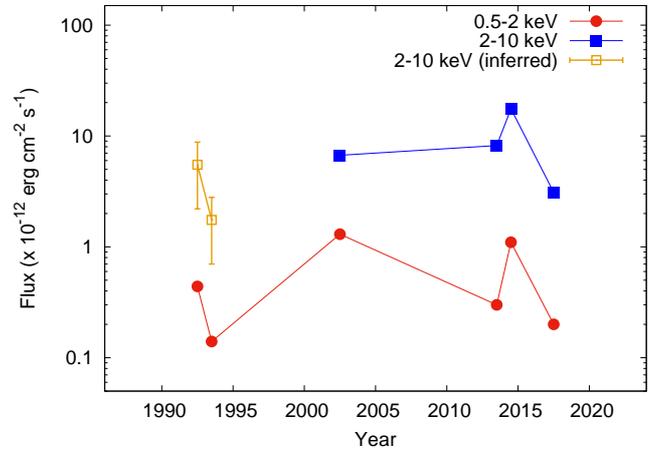}
		\caption{X-ray flux as a function of time collecting archival observations in the 0.5-2 and 2-10 keV band. The yellow squares are the inferred 2-10 keV flux considering the ROSAT fluxes and the two hardness ratios measured with XMM in 2000 and 2017. }
		\label{fig:lcurve_long}
	\end{figure}

	The flux obtained with spectral analysis was found to be roughly a factor $\approx$ 5 times higher than the one measured by \citet{pahari2017} 3 years before in July 2014. On the other hand, the estimated spectral parameters from the two datasets  were found be in good agreement, and the ratio of the 0.5--2 keV and the 2--10 keV fluxes does not show significant variations between the two epochs.  With this in mind, we also looked for other archival observations in the 0.5-10 keV band. We found that the source was observed once previously by XMM-Newton \citep{cappi2006} and by Suzaku  \citep{kawamuro2016} in 2002 and 2013, respectively. The  reported flux values indicate that only the first observation, performed more that ten years before the other three, presents a significant difference in the 0.5--2 / 2--10 keV flux ratio compared to the previous results ($\approx$ 0.2 vs $\approx$ 0.05). This is mainly due to a significantly lower absorption seen in the 2002 epoch.  Past observations with the ROSAT satellite in the soft band also showed significant variations \citep{polletta1996}, consistent with those observed in the recent years. The long term light curve in two bands is plotted in Fig. \ref{fig:lcurve_long} and shows how the X-ray emission changes significantly over the $\approx$ 30 years of observations, and in this last observation the source was at one of it's minimum levels. AGNs are known to be variable on long timescales. Such variations can be due to obscuring events, which change the spectrum, especially at low X-ray energies, or intrinsic variations of the power-law normalization. Given that the latter cause does not affect the slope of the spectrum (and therefore the spectral hardness), this suggests that at least the variations observed in the last 3 years are intrinsic and and originating in the immediate environment of the primary X-ray source.

	\subsection{Timing Analysis}

\begin{figure}
		\includegraphics[width=0.52\textwidth]{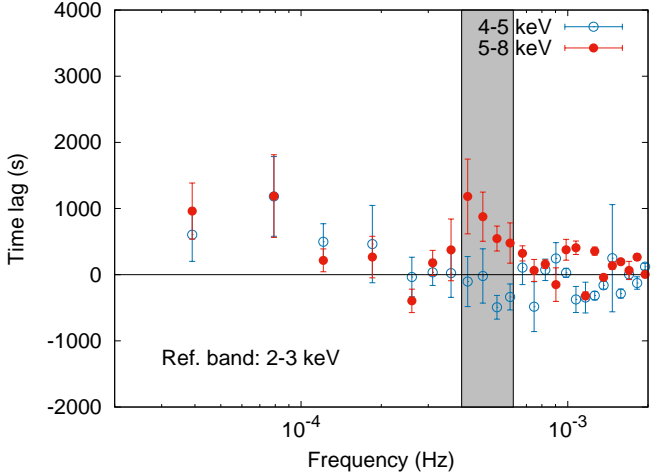}
		\caption{Time lags computed between the 2--3 keV band and the 4--5 (blue open circles) and 5--8 (red filled circles) keV band. Only the latter shows a lag between 4 and 6  $\times 10^{-4}$ Hz (highlighted in grey). }
		\label{fig:lag_frequency}
	\end{figure}

	In order to study the variability properties of this data-set we followed the procedures described in \citet{uttley2014}. To do this we used the open software Stingray \citep{Huppenkothen2019}.  In particular we computed the Fourier frequency dependent time lags with respect to  the continuum-dominated reference band of $2-3$ keV for the two subject bands: $4-5$ keV (Fig.~\ref{fig:lag_frequency}, blue open circles) and $5-8$ keV (Fig.~\ref{fig:lag_frequency}, red filled circles). In both cases, for lag-frequency spectra for frequencies lower than $10^{-5}$ Hz, we find a hint of a positive hard lag, possibly due to propagating accretion rate fluctuations. At higher frequencies instead we see a positive lag only between the 2--3 keV band and the 5--8 keV band, as expected for disc reverberation. 
	
	Following all previous papers on lag measurement, we have not corrected the data for possible red noise leakage distortion \citep[See e.g.][]{jenkins} which, if there is significant variability power on timescales longer than those observed here, might affect the lags. However there is no large scale trend obvious over the duration of the observed light curve ($\approx 100$ ks) so red noise leakage may not be large. Moreover we notice that based on the optically determined black hole mass, we do not expect large amplitude variability on timescales longer than those observed here \citep{mchardy2006}. 
	
	From our analysis of the time-averaged spectrum, we select the $2-3$ keV band as our reference direct continuum band for the calculation of lags. 
	Here a positive lag means that the harder subject band (i.e. $4-5$ or $5-8$ keV) lags the softer reference band ($2-3$ keV). In the $[4-6] \times 10^{-4}$ Hz frequency range, the $4-5$ keV band, which is dominated by the continuum shows a slightly negative lag, but is consistent with no lag (Fig.~\ref{fig:lag_frequency}).
	A clear positive lag is, however, visible in the same frequency range for the $5-8$ keV band that contains the iron line.
	
\begin{figure}
		\centering
		\includegraphics[width=0.5\textwidth]{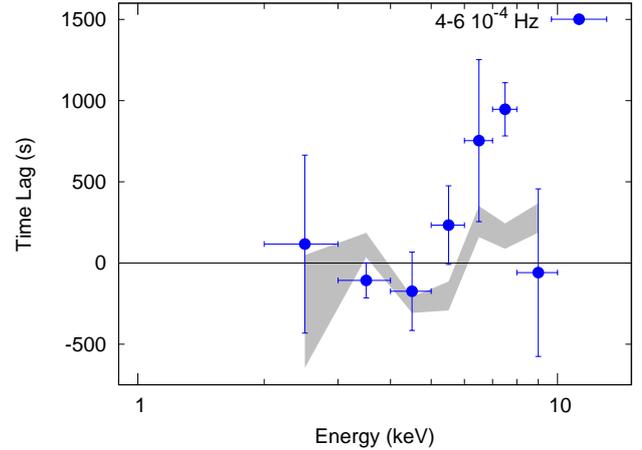}
		\caption{Time lag as a function of the energy. Lags were computed between 4--6 $\times 10^{-4}$ Hz. An excess around 7 keV is clear. Grey area are the lag vs energy computed between 0.9-1.5$\times$ 10$^{-3}$ Hz. }
		\label{fig:lag_energy}
	\end{figure}

	For the same $[4-6]\times 10^{-4}$ Hz frequency range we measure the energy dependence of the lags (Fig.~\ref{fig:lag_energy}). 
	To increase the S/N, we chose as reference band the larger 2-10 keV band, excluding the band relative to which the lag was computed. The lags are consistent with $0$ apart from a feature at $\sim 7$ keV, 
	further indicating the presence of a reverberation lag. {The lag vs energy dependence was also computed at higher frequencies (0.9-1.5$\times$ 10$^{-3}$ Hz, where only marginal evidence of a lag was seen). The grey curve in Fig\ref{fig:lag_energy}, shows that for this higher frequency range there is no significant trend which could suggest the presence of reverberation.}

	In order to quantify our result we also fitted the lag energy spectrum, with a Gaussian line. The best fitted model has a line center at $7.02^{+0.6}_{-0.22}$ keV with  $\sigma=0.7^{+0.3}_{-0.6}$ keV. 
	
	\section{Discussion}
	\label{sec:discussion}
	
	We have analysed the X-ray spectral-timing properties of the LLAGN NGC 5273 with a 90 ks continuous pointing. We detected for the first time in this source a signature of X-ray reverberation from an accretion disc in the time lags (Figs. \ref{fig:lag_frequency} and   \ref{fig:lag_energy} ).  On the other hand, we find only marginally significant evidence for a broad iron line in the time-averaged spectrum,
	such as is expected to arise when the inner edge of the accretion disc is very close to the black hole  \citep{ross-fabian1993,garcia2013}.  
	
	In order to shed light on this inconsistency we compared the observed lag with that measured in other AGN \citep{demarco2013,kara2016}. For consistency of band selection
	we therefore computed the lag between the 3-4 keV and 5-7 keV bands over the frequency range 4-6 $\times10^{-4}$ Hz, 
	finding a lag of $705 \pm 198$ s. Assuming a mass for the black-hole of  $4.7 \times 10^6$ M$_\odot$ \citep{bentz2014}, our measurement is compatible with the mass scaling relationships from \citep{demarco2013,kara2016} for both lag amplitude and lag frequency.  Even though such correlations seem to indicate a common geometry between Seyfert galaxies, they contain substantial scatter, showing that other effects are in play \citep{kara2016}.  
	
	A well known issue in time-lag measurements is ``dilution''. Both the direct and reflected components will be present in both the reference band and the channel of interest. Therefore the measured lag will differ from the intrinsic lag depending on the relative contribution between these two components. 
	
	A light crossing time of 700s corresponds to a distance of $\approx 20$ gravitational radii ($R_G$)  for a black-hole mass of $\approx 5 \times 10^6 M_\odot$. {Even though it has been shown the X-ray lag cannot be simply converted into light travel distance} this value suggests that the disc is truncated.
	It is perhaps reasonable to expect this source to have a truncated disc, since it has an accretion rate of $\sim 1$ percent of the Eddington limit \citep{pahari2017}, which is lower than the accretion rates of most other well known X-ray bright AGN in which broad iron lines have been found \citep{kara2016}. The location of the inner edge of the accretion disc as a function of accretion rate is not well measured, but there is evidence that the edge moves outwards as the accretion rate decreases \citep{cabanac}. 
	
	{Although the detection of a lag in the timing analysis is clear, the detection, or lack of detection, of a broad iron line in the time-averaged energy spectrum is less clear. The simple fit of
		\textsc{zpcf} $\times$ \textsc{tbabs} $\times$ (\textsc{pw} + \textsc{xillver}), which provides a power law and a narrow iron line, is an acceptable fit ($\chi^{2}$ of 180/163). As there are no clear residuals showing a broad line above the continuum, if we include a relativistic component (adding \textsc{relxillp}), the model prefers a solution with a very broad line and so it concludes that the inner edge of the disc must be at the ISCO.  This fit is not significantly better ($\chi^{2}$ of 175/160) than the one without a relativistic component (see Section 3.1).  Given the relative faintness of NGC~5273 then from the time-averaged spectrum we cannot conclude a great deal about any broad iron line except to say that it is not strong. A truncated disc may therefore provide an explanation of the the lack of a strong iron line in the energy spectrum, even though F-test reveals that a fit with a non-truncated disc is preferred with 2.8$\sigma$ confidence. However, any alternative fit with a truncated disc would also require a smaller reflection fraction in order to accommodate the correspondingly narrower relativistic iron line component. The extra dilution resulting from this lower reflection fraction would increase the inferred intrinsic light crossing lag.}

	In this context it is interesting to note that the largest lag measured here ($705 \pm 198$ s) is actually almost the same (1037 $\pm$ 455 s; \citealt{zogbhi2013}) as that measured in the higher mass MGC-5-23-16 ($8\times 10^7$ $M_\odot$); \citealt{ponti2012}).  From X-ray spectral analysis, the good agreement between the lags can be partially explained by the significantly lower reflection fraction found in MGC-5-23-16 (R=0.3; \citealt{zogbhi2017}). The dilution effect for MGC-5-23-16 will be higher than for NGC~5273, explaining why the observed lags in the two AGN are comparable. Moreover, the different reflection fractions suggest that the two sources have different disc configurations. Given that the inferred truncation radius of MGC-5-23-16 is of the order $\approx 50 R_G$,  these results seem to point towards a self-consistent picture.  
	
	The lag vs energy spectrum can also be used to extract physical information on the properties on the accretion flow. The energy dependence of the lag  shows a gradual increase before a sharp drop at 8 keV. Such a trend is in good agreement with the expectation from disc reverberation \citep{cackett2014,chainakun2019,ingram2019}.  Moreover the energy at which the lag peaks does not correspond  to the energy at which the iron line peaks in the energy spectrum (7 keV against 6.4 keV).  This was already observed for NGC 4151 by \citep{zoghbi2012}, and it is due to the different origin of the broad and narrow components in the time-averaged spectrum. While the first is explained in terms of an accretion disc very close to the central black-hole (giving origin also to the measured lags), the latter is believed to originate from a more distant and cold material \citep{zoghbi2012,zogbhi2019}.

	The value at which the lag vs energy drops is known to be directly related to the inclination of the disc \citep{cackett2014}. In particular, the more edge-on the source is, the higher the Doppler-shifted contribution will be, and therefore the higher the energy of the drop. It is possible to show that if the peak of the lag is after 7 keV, the inclination is higher than 45 $^\circ$ \citep[see Fig. 13 in ][]{cackett2014}. The high inclination seems also to be in good agreement with the general orientation-dependent picture of Seyfert classification. According to that scenario Seyfert galaxies which show mainly narrow lines (e.g. Sy 1.5 or 1.8) have a higher inclination (and therefore obscuration) than standard Seyfert galaxies. Of course other parameters could also affect the appearance of broader lines, originating from closer regions to the central black-hole (e.g. size of broad-line region or optical depth of the obscuring material). We note that a greater sample of spectral-timing measurements will permit in future to quantify the actual effect of the inferred accretion disc inclination on the optical lines.

	\section{Conclusions}
	
	We present a detailed characterization of the X-ray variability of the Seyfert 1.5 galaxy NGC 5273 from a 90 ks XMM-Newton observation. The source shows significant variability in the observations down to timescales of 1000 s. Applying cross-spectral analysis we found that the emission between 5 and 7 keV lags the rest of the continuum by $\approx$ 1000s. Given also the presence of a significant emission line in the X-ray spectrum at 6.4 keV we interpret this delay as the result of reverberation due to the illumination of the disc by X-rays from a hot central corona. From the shape of the lag-energy spectrum we found that the inclination of the source should be $\gtrsim 45^\circ$, which is in good agreement  with the classification of the source. The result presented here is a further confirmation of how powerful timing studies can be for probing the geometry of AGNs even in lower luminosity sources, which have not previously been explored with similar techniques.  Given the long term variability displayed by the source, new observations done at different flux levels could explore changes in geometry.
	
	\section{Acknowledgments}
	
	F.V. would like to thank Michiel van der Klis for the useful discussion and comments on the timing analysis of this dataset. IMcH and F.V. thank STFC for support under grant ST/M001326/1. MP acknowledges Royal Society-SERB Newton International Fellowship support funded jointly by the Royal So-
	ciety, UK and the Science and Engineering Board of India
	(SERB) through Newton-Bhabha Fund.
	
	
	
	
	
	\bibliographystyle{mnras}
	\bibliography{bibtex_agn} 

	\bsp	
	\label{lastpage}
\end{document}